%% file: main.tex
\documentclass[conference]{IEEEtran}
\IEEEoverridecommandlockouts
\usepackage{soul}

\usepackage{cite}
\usepackage{amsmath,amssymb,amsfonts}
\usepackage{algorithmic}
\usepackage{graphicx}
\usepackage{textcomp}
\usepackage{xcolor}
\usepackage{comment}
\def\BibTeX{{\rm B\kern-.05em{\sc i\kern-.025em b}\kern-.08em
    T\kern-.1667em\lower.7ex\hbox{E}\kern-.125emX}}
\begin{document}

\title{
Testing the data framework for an AI algorithm in preparation for high data rate X-ray facilities  
}

\author{\IEEEauthorblockN{Hongwei Chen$^{1, 2, 4*}$, Sathya R. Chitturi$^{1, 2, 3*}$, Rajan Plumley$^{1, 2, 5*}$, Lingjia Shen$^{1, 2}$, Nathan C. Drucker$^{1, 2, 6}$,\\ Nicolas Burdet$^{1, 2}$, Cheng Peng$^{2}$, Sougata Mardanya$^{7}$, Daniel Ratner$^{1}$, Aashwin Mishra$^{1}$, Chun Hong Yoon$^{1}$,\\  
Sanghoon Song$^{1}$, Matthieu Chollet$^{1}$, Gilberto Fabbris$^{8}$,
Mike Dunne$^{1}$, Silke Nelson$^{1}$, Mingda Li$^{9}$,\\ Aaron Lindenberg$^{2, 3}$, Chunjing Jia$^{2}$, Youssef Nashed$^{1}$,  Arun Bansil$^{4}$, Sugata Chowdhury$^{7}$,\\ Adrian E. Feiguin$^{4}$, Joshua J. Turner$^{1, 2}$, Jana B. Thayer$^{1}$}
\IEEEauthorblockA{\textit{$^{1}$Linac Coherent Light Source, SLAC National Accelerator Laboratory, Menlo Park, USA} \\
\textit{$^{2}$Stanford Institute for Materials and Energy Sciences, Stanford University, Stanford, USA}\\
\textit{$^{3}$Department of Materials Science and Engineering, Stanford University, Stanford, USA}\\
\textit{$^{4}$Department of Physics, Northeastern University, Boston, USA}\\
\textit{$^{5}$Department of Physics, Carnegie Mellon University, Pittsburgh, USA}\\
\textit{$^{6}$School of Engineering and Applied Sciences, Harvard University, Cambridge, USA}\\
\textit{$^{7}$Department of Physics and Astrophysics, Howard University, Washington, USA}\\
\textit{$^{8}$Advanced Photon Source, Argonne National Laboratory, Argonne, USA}\\
\textit{$^{9}$Department of Nuclear Science \& Engineering, The Massachusetts Institute of Technology, Cambridge, USA}\\
}
}

\maketitle
\begingroup
\renewcommand\thefootnote{*}
\footnotetext{Authors Contributed Equally}
\endgroup

\begin{abstract}
The advent of next-generation X-ray free electron lasers will be capable of delivering X-rays at a repetition rate approaching 1 MHz continuously. This will require the development of data systems to handle experiments at these type of facilities, especially for high throughput applications, such as femtosecond X-ray crystallography and X-ray photon fluctuation spectroscopy. Here, we demonstrate a framework which captures single shot X-ray data at the LCLS and implements a machine-learning algorithm to automatically extract the contrast parameter from the collected data.  We measure the time required to return the results and assess the feasibility of using this framework at high data volume. We use this experiment to determine the feasibility of solutions for `live' data analysis at the MHz repetition rate.

\end{abstract}

\begin{IEEEkeywords}
LCLS-II, X-ray, Machine Learning, Experimental Design, High Performance Computing
\end{IEEEkeywords}

\section{Introduction}

X-ray Free Electron Laser (XFEL) light sources are scientific user facilities with the goal of investigating atomic scale processes at ultrafast, femtosecond (1 femtosecond~=~10$^{-15}$ seconds) time-scales. \cite{Decking2020, Prat2020, Kang2017, Ishikawa2012, Emma2010, RevModPhys.88.015007}.  In a typical XFEL experiment, researchers from around the world are given access to the facility for a limited amount of beamtime to perform experiments at one of its many specialized scientific instrument end-stations.  Awarded beamtime is highly sought-after, and it is not unusual for users to wait months or years to carry out their experiments. The experiments are often highly complex, involving a number of additional capabilities, such as advanced laser systems, cryogenics, precision motion, ultra-high vacuum, high magnetic or electric fields, computational support through both controls and analysis, and highly sensitive samples.  Furthermore, the measurements involving the incident X-ray laser radiation must be remotely controlled from outside the local laboratory setup for the safety of the experimentalists.  With a period of beamtime access typically lasting between 12-60 hours, it is crucial that tools are in place so that XFEL users are able to use their time efficiently to collect data and perform the experiments.

Since their inception, XFELs have attracted much attention due to the new fields of science which they enable  based on their high pulse intensity, short pulse duration, available X-ray energies, and coherent properties\cite{Vinko2012, Bernitt2012, doi:10.1126/science.1229663, Wernet2015,doi:10.1126/science.abd7213}. A new generation of these lasers is currently being operated, constructed or planned at several sites around the world. With the increase in repetition rate, entirely new experimental methods will be made possible. Furthermore, increasing the XFEL repetition rate will also address feasibility challenges for certain experiments which currently take heroic efforts to perform, such as photoemission spectroscopy \cite{damascelli-rmp-2003}, transient grating spectroscopy \cite{Svetina:19}, and resonant inelastic X-ray scattering \cite{RevModPhys.83.705}.

One such technique is X-ray Photon Fluctuation Spectroscopy (XPFS), a method whereby one is able to use temporally separated X-ray pulses to measure fluctuations of a system by probing changes at different ultrafast timescales\cite{Shen2021}. Although XPFS can provide an unprecedented level of information in condensed matter systems, the data rates for a typical experiment at the next-generation machines will be intractable for realistic fast feedback during an experiment. This hindrance is due to the fact that $\sim$10$^{6}$ images on multiple mega-pixel detectors will be collected per second and analyzed `on the fly', where the analysis must keep up with the input data rate to extract the full value from the experimental time. This volume of data is a dramatic increase over the current XFEL data rate of hundreds or thousands of images per second.

In this paper, we address this problem of the data rate posed by an increase of XFEL repetition rates by using the LCLS-I data pipeline to run an experiment on a prototypical XPFS experiment and assess the scalability of our analysis methods. We measure scattered photons forming a speckle pattern from a Van der Waals antiferromagnet \cite{Burch2018}. The scattered photons are measured through the Data Acquisition (DAQ) system at the LCLS, reduced using a suite of tools called small-data tools, and analyzed using a new machine-learning algorithm running on a GPU system \cite{Sathya}. We benchmark the performance of the algorithm running within the LCLS-I analysis pipeline to evaluate its suitability for online data reduction and to decipher potential barriers for efficient use of the data systems at the data rates of the emerging generation of XFEL facilities such as LCLS-II, the European XFEL, and the SHINE facility --- up to the maximum rate of a continuous MHz data stream. We use these results to project the needs for capturing similar experimental data in the near future, and we anticipate how to include Bayesian optimization in this process to inform real-time steering of the experiment based on comparisons to theory and prior data. These results will be important for on-the-fly decision making and eventually, for experimental control at sophisticated XFEL beamlines.

\section{Motivation}
    
There are many applications where the method described here can be implemented. In general, these applications rely on the unique properties of XFELs but may also be extensible to the larger number of 4th generation synchrotron facilities as well. We focus here on the specific application of XPFS, where the temporal characteristics of the method match the natural timescale of thermal or quantum fluctuations of the different degrees of freedom in the system being studied. This has been shown by the progress made by XPFS in the field of topological magnets, but has also been applied to other systems of interest as well, such as the behavior of soft, disordered matter like colloidal nanoparticles \cite{sun-prl-2021}, and the exotic properties of water \cite{perakis-natcomm-2018}. In a series of studies \cite{seaberg-2017-prl,esposito-2020-apl,seaberg-prr, assefa-rsi-2022}, it was demonstrated that magnetic skyrmions -- spin textures with important topological properties for future electronic devices \cite{Fert2017} -- fluctuate in a distinct way on nanosecond timescales. Moreover, these fluctuations bare close similarity with the exotic physics observed in other seemingly disconnected systems, e.g. high-temperature superconductors \cite{seaberg-prr}. The discovery of these hidden connections highlights the power of XPFS, and more importantly, urgently calls for its implementation at the next-generation high-repetition rate XFEL facilities.  

Besides topological magnetism, many other areas within the condensed matter community are ripe for the implementation of XPFS, such as the potential for addressing quantum criticality. Unlike classical thermodynamical phase transitions, a quantum phase transition (QPT) occurs at absolute zero and is triggered by a non-thermal parameter \cite{sachdev-2001-book}. It is solely governed by Heisenberg's uncertainty principle, i.e. quantum fluctuations. This novel type of physics is thought to be important for unconventional superconductivity \cite{Broun2008}. However, the exact role of quantum fluctuations in materials of this class remains highly debated. One advantage of XPFS is that it can offer a direct evaluation of quantum order parameter fluctuations covering a broad time window ranging from the femtosecond level to hundreds of nanoseconds. The study of the dynamics of
the order parameter, for instance, could be explored far
away from to the regions in close vicinity to the quantum critical point. This capability of accessing fluctuations on these distinct timescales and over these ranges is not captured by any other current X-ray techniques.

\section{Application: XPFS}
XPFS relies on carefully comparing sequential scattering patterns of outgoing X-ray photons from the sample after it is illuminated by the incident XFEL beam. This process is akin to taking a series of photographs in very rapid succession, and adding them together while studying the ``fuzziness" of the image, to extract out valuable dynamics occurring in the system.  The primary quantitative metric used for this comparison is the contrast of the scattering intensity pattern collected by the X-ray detector.  Since the resultant scattering intensity or ``speckle"  pattern from the sample is directly related to its microscopic structure, one is able to draw conclusions about dynamical changes in the sample by monitoring the change in contrast \cite{SUTTON2008657, shpyrko2007direct}.  This method of analyzing changes in the sample is incredibly powerful for ultrafast physics experiments because its temporal sensitivity is only limited by the X-ray pulse separation time and not the readout time of the detector \cite{gutt2009measuring}.

\begin{figure*}[t]
            \centering
            \fbox{\includegraphics[width=0.95\textwidth]{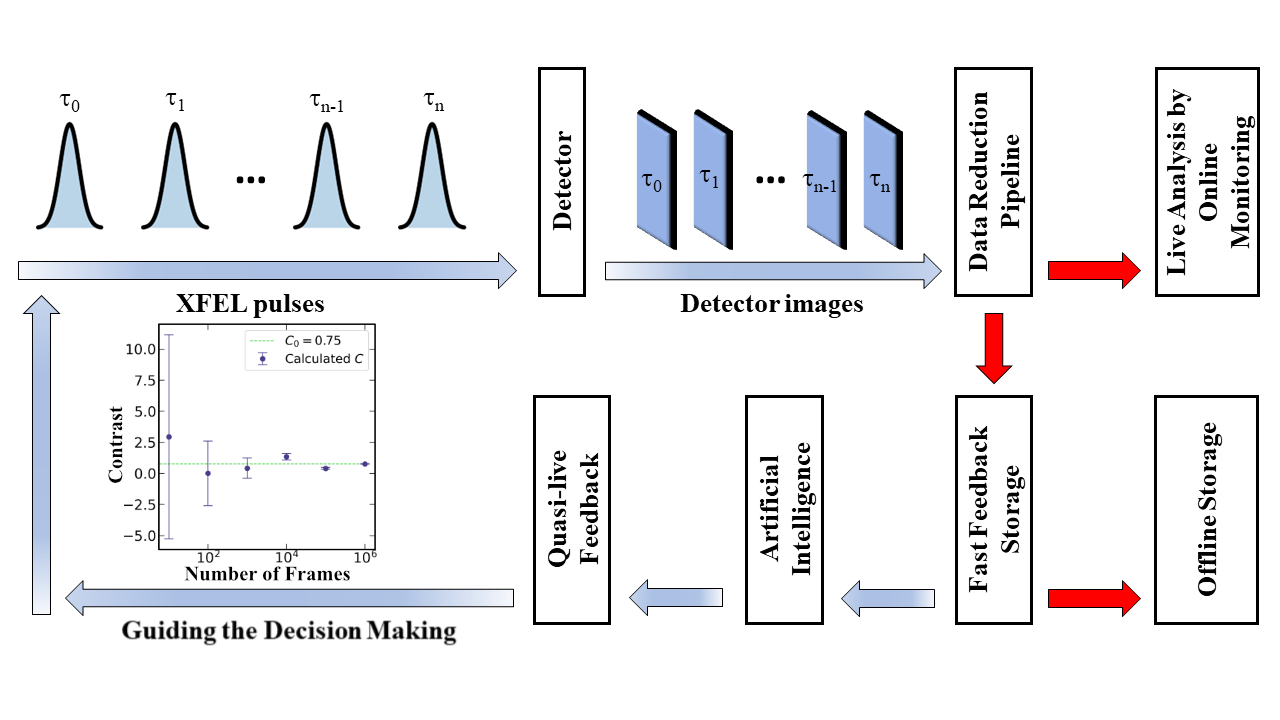}}
            \caption{Schematic of the analysis pipeline specific to our experiments described here for XPFS measurements. A series of X-ray pulses are measured at the LCLS after scattering from a sample and recorded on large-area detectors which is fed through both the Data Reduction Pipeline (DRP) and the Fast Feedback (FFB) storage systems. An artificial intelligence algorithm was used to analyze the ultra-fast scattering patterns and return the contrast parameter. Quick estimation of the contrast is crucial for understanding how the parameters of the experiment affect the signal to noise ratio as they are changed.  The plot shows calculated contrast vs. number of frames collected.  Error bars were calculated according to Eq.~\eqref{eqn:negative_binomial}.  Photon statistics were simulated using a classical Monte Carlo simulation on a 64x64 pixel region of interest.  The average count rate $\bar{k}$ ranging between 0.001-0.002 photons per speckle is well into the ``photon-hungry regime'' and is comparable to the count rate of the XPFS experiment described in this article.  The green horizontal line indicates the actual contrast value $C_0$.  Contrast values outside $[0,1]$ are unphysical.  It can take several minutes or hours to aggregate enough statistics for accurate contrast estimation at current XFEL repetition rates, however, this time will be reduced to seconds with the introduction of high repetition-rate facilities such as the LCLS-II.}
            \label{fig:lcls_data_system}
\end{figure*}

However, accurate contrast determination requires a large number of measurements as well as a spatially precise method for mapping the photon hits in the detector. This becomes especially important in the so-called photon-hungry regime, where the next-generation XFEL facilities benefit strongly. The challenge associated with mapping the single photon distribution is due to the problem of X-ray photons exciting a charge cloud in the detector sensor that is comparable to or larger than a single pixel \cite{Shen2021}. %
This ``bleeding'' of charge into adjacent pixels creates characteristic artifacts where the signal associated with a single photon is spread across multiple pixels, creating a degree of uncertainty in the scattering patterns.  This is dealt with using a greedy guess algorithm for reconstructing the photons from their charge clouds\cite{Burdet2021, sun2020accurate}.  The statistical aspect of the data collection can be understood, in the sparse photon limit, by the following equation, where the standard deviation $\sigma_c$ for contrast $C_0$ scales with the average photons/speckle $\bar{k}$, number of speckles $N_\text{speckle}$, and number of image frames collected $N_\text{frame}$ as \cite{sun2020accurate}:
\begin{equation}
    \frac{\sigma_c}{C_0} \simeq \frac{1}{C_0 \bar{k}}\left[\frac{2(1+C_0)}{N_{\text{speckle}}N_{\text{frame}}}\right]^{1/2}
\label{eqn:negative_binomial}
\end{equation}
Both criteria for statistical volume and accurately reconstructed photon-placement must be satisfied in order for contrast to be reliably determined.  In the case of an XFEL experiment where users are relying on analysis feedback in order to make critical and timely decisions related to what direction to take an experiment, these challenges must be overcome by a solution which provides real-time results so that beamtime can be used effectively. \\

\section{Experimental}
In the XFEL experiment used to benchmark this study, two colinear XFEL pulses with photon energy of 8.3 keV and time-separation of $20$ femtoseconds were incident on the sample at a rate of 120 times per second\cite{Decker2022, sun2019compact}.  This was generated by both a self-seeding scheme, as well as the slotted foil at LCLS. The resultant scattering patterns were collected by four EPIX100a detectors running at the same rate. For this experiment, the raw detector data is of dimensionality $N_{speckle}\times4\times773\times709$. Here $773\times709$ is the number of speckles in a given frame for one of the four detectors. In addition a data payload, including images and waveforms from several experiment diagnostics, are immediately saved to disk and simultaneously processed online in the LCLS-I Fast Feedback (FFB) processing farm.  The FFB is a standard High Performance Computing (HPC) system which offers dedicated resources to the running experiment in order to provide quasi-real-time ($< 1$ minute) feedback about the quality of the acquired data.  The FFB layer provides the first persistent storage layer in the data flow chain and also offers the first processing layer where experimenters can access the full dataset.  Ultimately, the algorithms evaluated in the LCLS-I FFB may be run online in the LCLS-II Data Reduction Pipeline (DRP). The DRP is a real-time analysis pipeline and hence must keep up with the input rate, currently to process data within the ms timeframe. The DRP allows one to plug in experiment-specific algorithms that operate on the streaming data. So, it is possible in LCLS-II to process the data and decide whether to keep or toss the event before writing to disk. We describe the full data systems in more detail in the following section.

\section{LCLS Data System}
Before detailing the tests, we give an overview of the LCLS data systems for a general background. For more information, see the paper by J. Thayer et al.\,\cite{thayer-jac-2016}. A schematic of this process relevant to the experiments presented in this paper is shown in Fig.~\ref{fig:lcls_data_system}.

\subsection{Overview}
The LCLS-I Data System is designed to acquire and to reliably transport shot-by-shot data at a peak throughput of 5 GB/s to the offline data storage where experimental data and the relevant metadata are archived and made available for user analysis. The LCLS data systems acquire data at the 120 Hz repetition rate of the LCLS light source, write multiple GB/s to a data cache, transfer data to offline storage and tape, and provide analysis software for the timely processing of this large and complex data set. A data management system and electronic logbook are available to users for managing the experiment and associated data files.  

Most devices are operated at the LCLS trigger rate of 120 Hz.  However, the DAQ is also able to acquire data from devices that read out at different frame rates (e.g., 120Hz, 30Hz, and 10 Hz) and associate the data with the appropriate fiducial in the event record. The shot-by-shot data are bundled into a run, defined as the set of data recorded for some period of time with a constant configuration. The DAQ system performs real-time assembly of the data from all devices into one object, called an event, tagged with the fiducial from the timing system and a UNIX timestamp and appends these event data to a file in extended Tagged Container (XTC) format \cite{aubert-jac-2013}. In the case of a dropped packet or missing data contribution, the metadata associated with the event is annotated appropriately. The data are cached in SSDs, and the transfer to Fast Feedback (FFB) nodes is initiated immediately when a run is started. The average size of a dataset ranges from 1 TB to tens of PB, and the data cache is sized to hold 1 shift (12 hours) of reduced data at the maximum rate. The fast feedback nodes can store 100-200 TB of data while awaiting transfer to permanent offline storage. The 100 - 200 TB depends on the maximum write rate of raw data, which is about 1 - 2 GB/s in total.  Assuming we record all data 100\% of the time during a 12 hour shift, 1 GB/s continuous will produce about 43 TB/day. Data can be accessed from disk, and custom analyses may be run on the fast feedback queues in each experimental hall providing quasi-real-time feedback within about 5 minutes of data acquisition.

\subsection{LCLS-II data systems}
The LCLS-II upgrade will increase the repetition rate from 120 Hz to a staggering rate of 1 MHz.  Coupled with the adoption of ultra-high repetition rate imaging detectors, there will be a significant increase in the data throughput from today’s maximum of 5 GB/s to 200 GB/s at first light and reaching multiple TB/s by the end of the decade. The LCLS-II system will have a Weka-based file system consisting of 16 server nodes, each one with 8 NVMe SSDs. The rate scales with the number of servers, so the write-rate can be improved by increasing the number of servers. Each individual drive is capable of writing between 1-2 GB/s which matches the rate of the source data. When the file system is read and written at the same time, the read rate is around 120 GB/s and the write rate is around 70 GB/s. 

Corresponding advances in data handling and computing are required to manage the quality and quantity of the data in order to provide flexible and easy-to-use fast feedback to users in real-time, a challenge given the weekly turnaround of experiments and the number of new user groups. A new Data Reduction Pipeline, an online filtering and feature extraction computing layer, is required to remove non-essential information from the data, enhancing the quality of the data that are taken while simultaneously reducing the data volume. 
FFB provides the first persistent storage layer in the data flow chain and also offers the first processing layer where the users can access the full data set.  In those science cases where the full analysis demands several petaFLOPS of processing capabilities, the FFB will run a simplified version of the analysis: fast enough to run in the terascale and sophisticated enough to provide event-level information which can be used to drive the experiment.
In the LCLS-II data system, the FFB processing nodes are connected to the DRP through the InfiniBand data network. NVRAM technologies are used for the storage layer as opposed to spindle-based storage as these technologies excel at handling the high concurrency generated by the DAQ writers, the data movers, and the FFB analysis. LCLS-II data cache is currently 780 TB.  We estimate that about 1 PB/shift is adequate if we employ data reduction and take into account the typical duty rate of experiments.

The new LCLS-II Data System has been deployed to the TMO and RIXS instruments and will eventually be rolled out to all LCLS instruments.  For each instrument there is a dedicated infrastructure for reading out detectors. The Data Reduction Pipeline and Fast Feedback layers are shared by multiple instruments. The offline data center is also a shared resource and may be supplied by local resources at SLAC or by remote High Performance Computing (HPC) facilities such as NERSC \cite{thayer-jac-2016, Damiani-psana}.

\subsection{Data management, software, and tools}
LCLS has developed a powerful data management system, deployed to all instruments  that handles both the automatic workflows of the data through the various storage layers — such as long term data archiving — and user requests through a web portal, such as restoring data from tape. And it can be used by many users simultaneously. The data management architecture, which allows transparent integration of both local and external HEC facilities, is capable of automatic run processing based on triggers such as start run, end run, first file transferred, all files transferred, or manually triggered.  This capability enables the development of sophisticated and interconnected workflows that automatically process the data, then take subsequent steps based on the results. 

PSANA (Photon Science ANAlysis) is a software framework that is used to analyze data produced by the LCLS X-ray free electron laser. Beginning in 2011, the project is written primarily in C++ with some Python, and provides user interfaces in both C++ and Python although most users use the Python interface.  The same code can be run in real-time while data is being taken as well as offline, executing on many nodes/cores using MPI for parallelization. PSANA facilitates these essential features:  moving data from persistent storage to memory, transparently handling the perfectly parallel nature of LCLS data (for most experiments, each event can be processed independently), handling detector calibrations, and invoking science-specific algorithms.

SmallDataTools is a suite of code useful for analysis from the XTC data to smaller HDF5 files at several stages of analysis. It is most frequently run using the FFB layer operating on data as they are acquired to generate reduced data sets in a new format.  The first step is the generation of the ``small data" file, the colloquial name for run-based HDF5 files which contain data arrays where the first dimension is the number of events (shot-to-shot information retained).  These files are automatically generated by the data management system as each new run is acquired so that the data are available only a few minutes after the run has ended. It can also be run on request to explore the effect of different parameter values for the data extraction.  Processing of the area detector can also be performed, such as extracting a region of interest, azimuthal integration, photon counting, etc.

\section{Machine Learning System}

In this section, we describe the ML model used to perform the photon assignment task, which is a typical analysis at the LCLS. First we review the model architecture and training procedure and then we characterize the performance of the model when deployed within the automatic data acquisition and analysis pipeline running offline on a Nvidia GTX 1080Ti GPU. This experiment is expected to be comparable to live analysis using the FFB since the main difference is due to data transfer speeds and job allocation. 

\subsection{Model} %

A fast machine learning algorithm has recently been developed for the task of converting raw XPFS images to photon maps \cite{chitturi2022machine}. This is a critical function in generating the contrast values for a given measurement, which must be collected until reasonable statistics are captured. A typical plot of this is shown in the inset of Fig.~\ref{fig:lcls_data_system}.

\begin{figure}[t]
            \centering 
            \includegraphics[width=87mm]{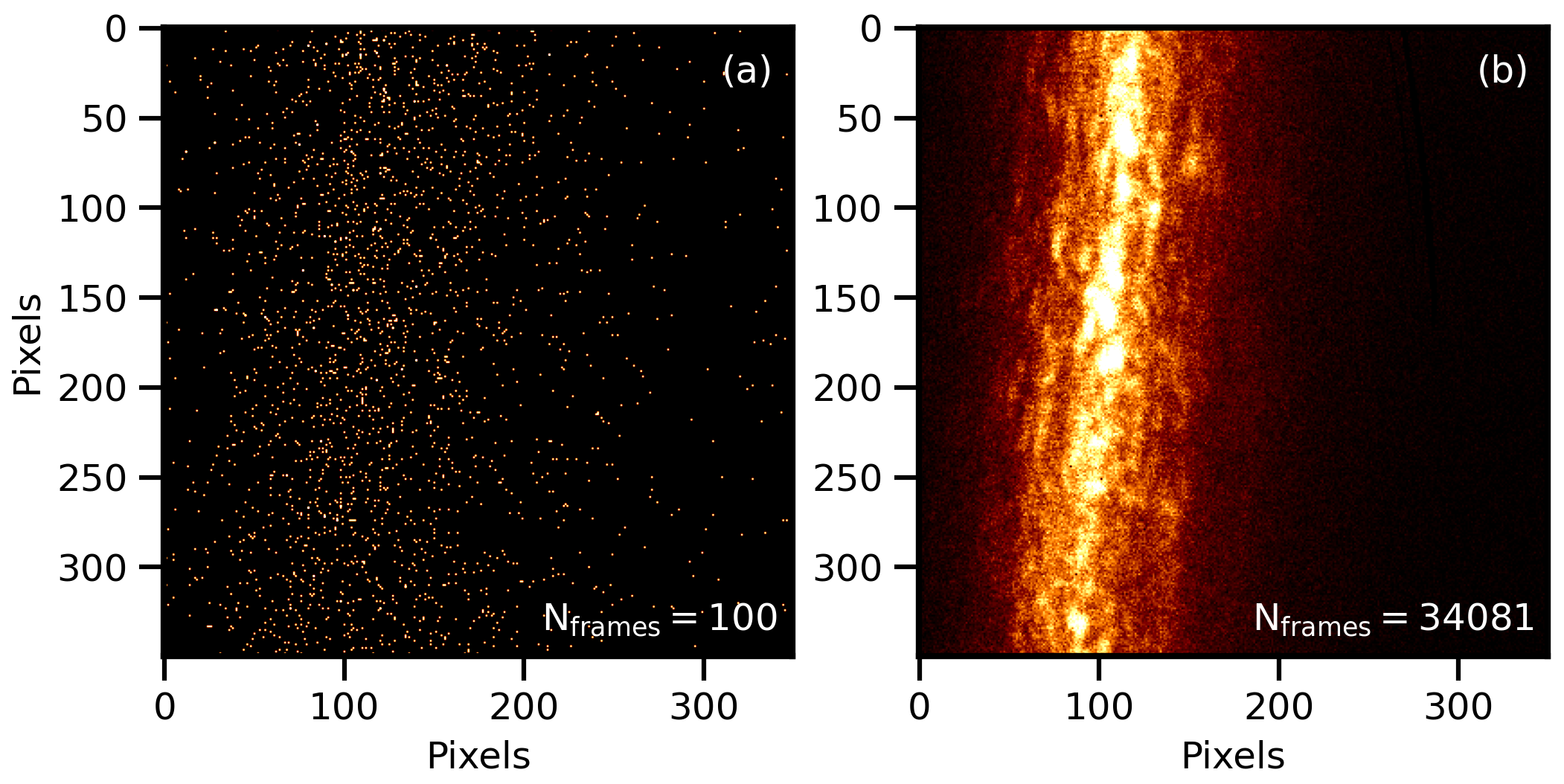}
            \caption{X-ray speckle pattern originating from the scattering from a single crystal NiPS$_3$ sample.  Photon hits on the EPIX100 detector are aggregated over $N_\text{frames}$ number of exposures.  (a)  100 exposures, or approximately 1 second of data collection at the LCLS.  (b)  34,081 exposures, which when acquired at a 120 Hz repetition rate, is equivalent to approximately 5 minutes of data collection time.}
\end{figure}

In this analysis, it is found that the machine learning model is able to outperform conventional methods in experiments with low contrast and experiments with high count rates \cite{chitturi2022machine}. Furthermore, the machine learning model is much faster than its classical counterpart, especially when parallelized on standard GPU hardware. Similar to Chitturi et al.~\cite{chitturi2022machine}, the present work also uses a U-net architecture \cite{ronneberger2015u} and frames the problem as a regression problem. Here we note that the U-net architecture was chosen as it is an established fully-convolutional architecture for image segmentation. In particular, our model choice of a convolutional neural network was inspired by the insight that the features of interest are naturally local -- i.e. charge sharing only occurs in nearby pixels. Furthermore, the fully-convolutional architecture allows for prediction on arbitrary sized regions of interest \cite{chitturi2022machine}. 

Here, the model is trained to predict the exponential of the photon map, instead of simply the photon map. The exponential function is used to address the highly sparse labels due to low X-ray count rates for the experiment. The exponential function is used to address the highly sparse labels due to low X-ray count rates and the statistics for the experiment. For example, due to the negative binomial statistics underlying these experiments, high photon events (for example, three or four photons per-pixel) are extremely rare relative to low photon events (for example, zero or one photon events). The exponential transform weights these high photon events at a larger rate relative to low photon events. This weighting teaches the model to learn to better predict high-photon events, which is crucial for good contrast estimation. In addition, we found that gradient descent is significantly smoother when transformed labels are used. The model is trained using an accurate simulator that is optimized for the EPIX100 detector, with its particular point spread function and noise characteristics \cite{sun2020accurate} and optimized using the Mean Squared Error (MSE) loss function between the predicted and ground-truth exponentiated photon map. Other loss functions such as categorical cross-entropy and negative log-likelihood were considered \cite{chitturi2022machine}, however we found that the MSE gave the best performance on a held-out validation set. The hyperparameters used for training are reported in Table \ref{tab:hyperparameters}. At test time, the logarithm operation is applied to the model output to retrieve the output photon map. For the purposes of this analysis, testing was done on tensors of shape $2\times128\times128$. Here, the $2$ represents two separate EPIX100 detectors and $128\times128$ is a typical image size for a cropped region of interest on the EPIX100 detector for LCLS-II experiments \cite{https://doi.org/10.1107/S1600577516010869}. 

\begin{table}[b]
\centering
\caption{Machine Learning Training Parameters}
\begin{tabular}{|l|l|l|l|l|}
\hline
Batch Size             &  128   \\ \hline
Learning Rate          &  0.01    \\ \hline
Optimizer              &  Adam    \\ \hline
Dataset Size           &  5000  \\ \hline
Dataset Split Fraction (Training/Validation)  &  0.1   \\  \hline
GPU Type               &  NVIDIA A100  \\ 
\hline
\end{tabular}
\label{tab:hyperparameters}
\end{table}

\subsection{Results: ML Deployment} %

\begin{figure}[t]
            \centering
            \includegraphics[width=80mm]{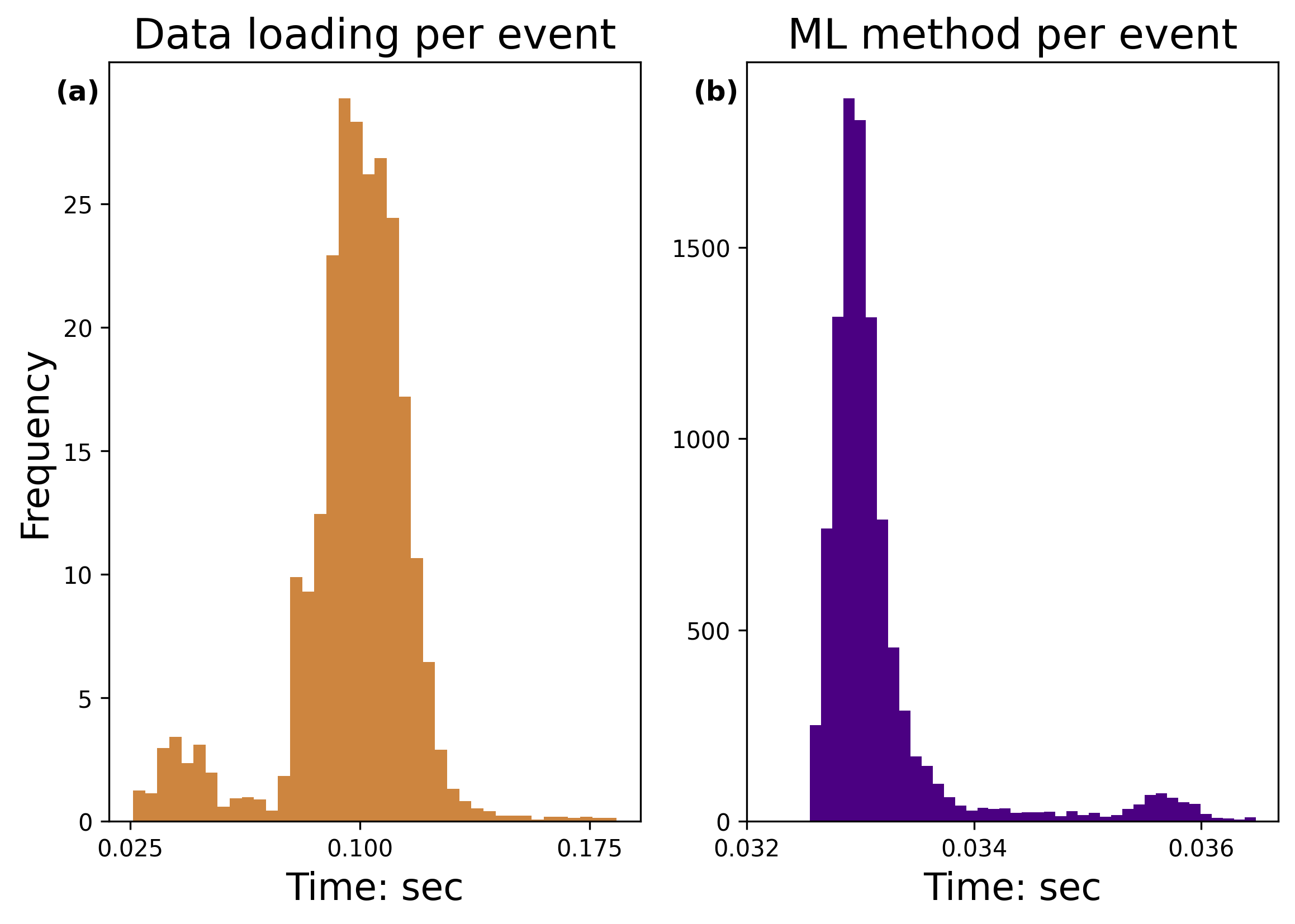}
            \caption{ (a) The distribution over time for converting the raw data from the XTC file format to the H5 file format on a CPU; (b) The distribution over time for processing each event using an ML model. Data are collected from 10,000 detector events for both graphs.}
            \label{fig:distribution}
\end{figure}

We characterize the performance of the ML model in two modes: streaming and batch configurations. In streaming, the ML model runs sequentially over individual frames as they are collected. In this scenario, the ML model makes predictions on individual events, data of shape 
$2\times128\times128$. For the streaming case, we report the distribution (Figure \ref{fig:distribution}) over timings for SmallDataTools to reduce the raw data and for the ML model to make a prediction for each event on a GPU, where the time spent on data transfer between GPU and CPU, and post-processing on CPU is included. In Figure \ref{fig:distribution} (a) and (b), most data fall into a small region around the mean while both curves have a small tail, which are probably caused by hardware. The distribution of data loading time is more dispersed compared to the time for the ML method. 

In contrast to streaming, where the model is run as individual events are streamed (Figure \ref{fig:distribution}), another approach is to wait and aggregate sufficient data before triggering a model prediction call (i.e. batch prediction). This approach has the advantage of much higher throughput (number of frames analyzed per second) as it exploits the parallelism of GPU architectures which are designed for operations on large tensors \cite{huyen2022designing}. However, the disadvantage of this approach is that the time to analyse one batch of data, under the assumption of that one batch is processed per unit time, will increase with batch size \cite{kirk2016programming} (i.e. higher latency). In general, deciding whether it is better to have low latency or high-throughput depends on the specific application and user needs. In this specific problem, it is interesting to note that the observable of interest, the contrast, is typically a property of a batch of images (not simply a single image). This is because a large number of frames are needed to be aggregated to have confidence in the statistics of the experiment (Equation \ref{eqn:negative_binomial}). 

\begin{figure}[t]
            \centering
            \includegraphics[width=70mm]{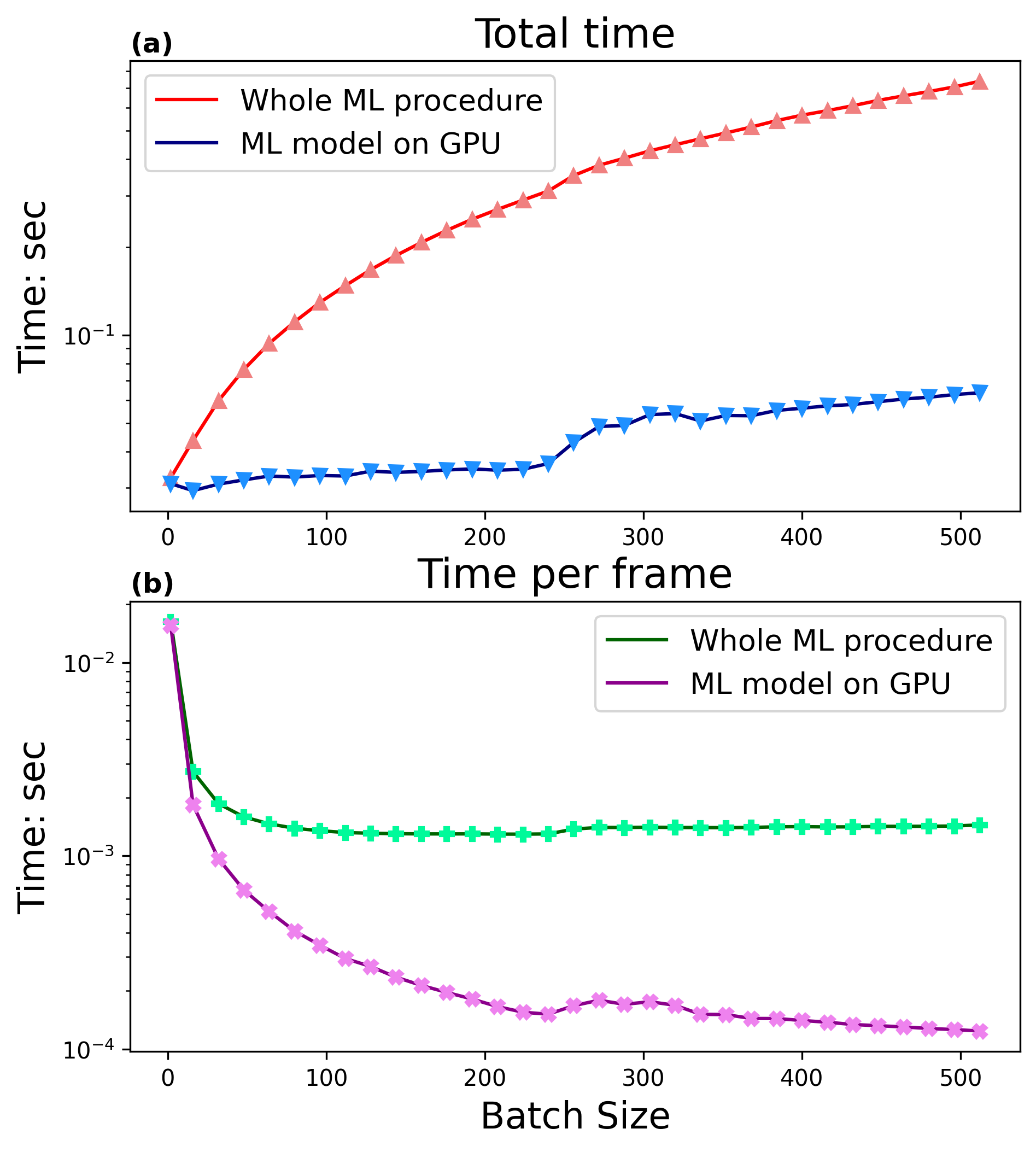}
            \caption{ (a) The relation between batch size and the total running time for the whole ML procedure, and the model running time on the GPU for one batch; (b) The relation between batch size and time-per-frame of the entire ML procedure, together with the model running time on the GPU. All the calculations are performed on the Intel(R) Xeon(R) CPU E5-2620 v4 CPU and Nvidia GTX 1080Ti GPU.}
            \label{fig:ml_batch_latency}
\end{figure}

This trade-off between latency and throughput is illustrated in Figure \ref{fig:ml_batch_latency}. Here, both the running time for the ML model on a GPU and the time for the entire ML procedure, including the running time on the GPU and data transfer, are shown. Figure \ref{fig:ml_batch_latency} (a) shows the total time needed for prediction of one batch as a function of batch size while Figure \ref{fig:ml_batch_latency} (b) illustrates the time-per-frame as a function of batch size. For very small batch size, the total time for the full ML procedure depends on the execution time on the GPU while in the case of the large batch size, it is dominated by the time on data transfer to the CPU. 

From this trade-off visualization in Figure \ref{fig:ml_batch_latency}, we can see that the prediction time per frame decreases sharply with batch size while the total time to make a prediction increases. In other words, we can gain a factor of approximately 10, but in principle, a larger batch size could have been used. Currently, we were limited by the total memory of the 1080Ti GPU -- so with access to more GPUs or better GPUs-- these speed gains can be even faster.  

\section{Experimental Design}
The study of materials through experiment requires exploring a high-dimensional space of parameters within a phase space constrained by time, sample availability, and cost. Typical sampling strategies blindly search through all possible combinations without leveraging the information being acquired, thus being unable to adaptively narrow the space of optimized solutions. The role of theoretical models (and computational simulations based on such theoretical models) in experiments is to narrow down regions of interest in a vast multi-dimensional phase space that include relevant XPFS experiments such as temperature, magnetic field, electric field and momentum transfer, $q$. However, while such models and simulations can enable scientists to more efficiently explore the high-dimensional parameter space present in these measurements, they can fall short because of their high computational cost or disconnect from real experiments. As a result, experiments at synchrotron or XFEL light sources have been guided by previous empirical experimental data and limited numerical modeling of system dynamics. 

To improve data acquisition efficiency in the experimental workflow, it is crucial to close the loop between theory and experiments in near real time. Having access to the contrast per batch discussed above for XPFS will allow for quasi-real-time analysis of theoretical models, enabling the user to update experimental conditions to maximize information gain from subsequent measurements. Experimental design frameworks such as Bayesian optimization (BO) can incorporate experimental and computational data to guide data acquisition based on provided `acquisition function,' for example to minimize the discrepancy between theoretical predictions and experimental results. Critically, to implement experimental design at an XFEL, the data analysis pipeline should be fully automated and capable of providing results on the same time scale as the batch acquisition time.

\section{Summary and Outlook}

In this work, we demonstrated that it is possible to implement an ML algorithm in the small-data tools at LCLS and run during the course of an active beam-time. We found that this implementation exhibits a trade-off between latency and throughput. Regarding the U-net model, one important future direction is to investigate whether reducing the number of layers and filters in the U-net could improve throughput and reduce latency, while preserving similar levels of predictive accuracy; here, complementary approaches such as model distillation and quantization could also be interesting avenues for future work \cite{polino2018model}. 

For the XPFS application specifically, using batch prediction is particularly favorable as the contrast calculation necessarily requires a large number of frames to obtain optimal counting statistics. However, the data transfer between GPU to CPU is quite slow for large batch sizes and greatly increases the latency of the system. An alternative approach could be to avoid this transfer entirely and instead perform the contrast calculation, per-batch, on the GPU. As long as the batch size is sufficiently large, this would give good enough statistics for live-monitoring. Of course, in the offline analysis, all the frames can be aggregated and analyzed, yielding even better statistics. 
   
Finally, we provide an outlook for future studies. If we make an assumption that the data transfer between GPU to CPU is not required and that contrast can be `streamed' on the GPU, then 1 GPU will achieve a rate of about $\sim$ 10 kHz. Note that this is with the use of a 1080Ti, but we could expect about $\sim$ 5x better performance with more advanced hardware in the near future. We conclude that a cluster of about 20 GPUs would get to the ultimate goal of a 1\,MHz analysis capability, with the worst case 100 GPUs with the current hardware benchmarked for this study. %
In general, the batch size needed for reasonable statistics will depend on the contrast level and the count rate of the particular experiment, but for some experiments, this could be in the 10s or 100s of thousands of frames. With developments in hardware and the ability to access even greater amounts of GPU memory, these requirements could be met, making this experimental capability at the next generation XFEL light sources within reach.

\section*{Acknowledgment}

The authors thank Yanwen Sun for discussions regarding the detector simulation. This work was supported by the U.S. Department of
Energy, Office of Science, Basic Energy Sciences under Award No. DE-SC0022216. J. J. Turner acknowledges support from the U.S. DOE, Office of Science, Basic Energy Sciences, Materials Sciences and Engineering Division through the Early Career Research Program, under Contract DE-AC02-76SF00515. The use of the Linac Coherent Light Source (LCLS), SLAC National Accelerator Laboratory, is supported by the DOE, Office of Science under contract DE-AC02-76SF00515. M.L. acknowledges support from U.S. Department of Energy (DOE), Office of Science, Basic Energy Sciences (BES), award No. DE-SC0021940. N.D. was supported by the U.S. Department of Energy, Office of Science, Office of Workforce Development for Teachers and Scientists, Office of Science Graduate Student Research (SCGSR) program, under contract number DE‐SC0014664.

\bibliographystyle{IEEEtran}
\input{ref.bbl}

\end{document}